# *"On the Road"* - Reflections on the Security of Vehicular Communication Systems

Panos Papadimitratos
Laboratory for Computer Communications and Applications (LCA)
Ecole Polytechnique Fédérale de Lausanne (EPFL)
Lausanne, Switzerland
`panos.papadimitratos@epfl.ch`

*Abstract*— Vehicular communication (VC) systems have recently drawn the attention of industry, authorities, and academia. A consensus on the need to secure VC systems and protect the privacy of their users led to concerted efforts to design security architectures. Interestingly, the results different project contributed thus far bear extensive similarities in terms of objectives and mechanisms. As a result, this appears to be an auspicious time for setting the corner-stone of trustworthy VC systems. Nonetheless, there is a considerable distance to cover till their deployment. This paper ponders on the road ahead. First, it presents a distillation of the state of the art, covering the perceived threat model, security requirements, and basic secure VC system components. Then, it dissects predominant assumptions and design choices and considers alternatives. Under the prism of what is necessary to render secure VC systems practical, and given possible non-technical influences, the paper attempts to chart the landscape towards the deployment of secure VC systems.

## I. INTRODUCTION

Over a number of years, Intelligent Transportation System (ITS) and related technologies have been deployed; for toll collection, fleet logistics and management, anti-theft protection, pay-as-you-go insurance, traffic information, and active road-side signs. Such systems, relying on different communication technologies, continue to evolve and proliferate. A new trend has recently emerged: the integration of *on-board computing units* (OBUs) and IEEE 802.11p radios for *vehicle to vehicle* (V2V) and *vehicle to road-side infrastructure* (V2I) communication. These *vehicular communication* (VC) systems enhance transportation safety and efficiency. V2V communication enables real-time safety applications, extending the driver's horizon, while both V2V and V2I communication enhance the versatility and effectiveness of applications for environmental awareness, VC system information dissemination, and in-vehicle entertainment.

However, the unique features of VC are a double-edged sword: a rich set of tools are offered to drivers and authorities, but a formidable set of abuses and attacks becomes possible. An attacker could "contaminate" large portions of the vehicular network with false information, announcing, for example, non-existent dangerous or congested road conditions, misleading drivers and causing traffic jams. Or, drivers could purchase software or hardware VC system "hacks," as they now often purchase police radar detectors or modify their cars for additional horse-power. Such VC system modifications could, for example, allow private vehicles to transmit messages as if they were an emergency vehicle (ambulance, police patrol, or road maintenance vehicle), have unsuspected drivers notified by their OBUs to slow down and yield, and this way offer fast movement even in traffic jams. From a different point of view, receivers deployed in a city center, at highway exits, or even in a celebrities neighborhood, could record transmissions from by-passing vehicles to later trace their location and infer private information about their passengers.

These exploits show that security is needed, especially because attacks are relatively easy to mount. First, VC rely on the widely adopted IEEE 802.11 wireless communication technology. Moreover, attackers could use any low-cost computing platform, such as PDAs, laptops, or WLAN access points (APs); for example, a wireless network operator, licensed to provide services unrelated to VC systems, could "tune" its APs to intercept VC traffic. Finally, VC equipment can be left unattended for long periods, increasing the likelihood of physical compromise. Overall, without security, VC systems could make anti-social and criminal behavior easier, in ways that would actually jeopardize the benefits of their deployment.

The awareness on the need to secure VC spurred a number of activities, with ongoing efforts, by *SeVeCom* [6] and IEEE 1609.2 [4], to design VC security architectures. Sec. II provides a condensed survey of the most recent understanding in terms of threats, requirements, and mechanisms. But with significant commonalities among results from different projects, *is the problem largely solved? Are there any research questions that remain to be addressed? Or, is it more or less clear which VC security architecture will be instantiated and deployed?*

In this paper, we reflect exactly on these questions, which would be tempting to answer in an affirmative manner. We revisit a number of commonly understood technical approaches and discuss alternative view points. Then, we consider non-technical factors that are likely to influence the deployment of secure VC systems. In spite of significant progress in terms of security, several issues remain to be addressed and system aspects to be crystalized towards deployment. At the same time, the development of security architectures can influence the design of VC systems. Even though surprising turns or obstacles can always be encountered, one can be optimistic that trustworthy VC systems will be deployed.

## II. SECURE VC SYSTEMS

### A. Adversary Model

VC system entities can be *correct* or *benign*, that is, comply with the implemented protocols, or deviate from the protocol definition being *faulty* or *adversaries*. Faults may not be malicious; for example, the communication module of a *node* (vehicle OBU or road-side unit (RSU)) may discard or delay messages or set packet fields to inappropriate values. But malicious behavior can result in a much larger set of faults. A detailed discussion of adversary models and aspects germane to VC systems is provided in [8]. Adversaries can be *internal*, with cryptographic keys and credentials to participate in the protocol(s) execution. Or, they can be *external*, but still able to influence the protocols by jamming communications and replaying messages of other nodes.

Even though the VC implementations will be proprietary, standards, needed for inter-operability, will provide extensive information on the VC protocol stack. Attackers will be able to build their own rogue protocols and modify the functionality of VC system nodes. If they obtain compromised cryptographic keys, for example, physically extracted from a node, they can act as internal adversaries. In fact, a node holding multiple such keys can appear as multiple nodes.

More generally, many adversarial nodes can be present. Often, they can act individually, but they might also act in collusion, coordinating their actions. Even if they do so, they should be unwilling to share their private keys and allow another node to fully impersonate them (and obtain, for example, their access rights). Over time, the number of adversaries can change, depending on the type of compromise as well as the defensive reaction of the system. It is reasonable to expect that at any point in time a small fraction of the network nodes are adversaries. At any time and location, only a few adversaries are likely to be physically present. This does not preclude a group of adversarial nodes surrounding a correct one, but such a situation should be rare.

A rather peculiar type of adversary is relevant to VC systems: an *input-controlling adversary*, which alters (sensory) inputs to VC protocols, rather than compromising the protocols. Such an adversary is weaker than an arbitrary internal adversary, because it cannot induce arbitrary behavior. But it will be often much easier to affect inputs, or compromise sensors or sensor-OBU connections, than compromising the OBU itself.

### B. Security Requirements

Without considering specific applications and protocols, a list of general security requirements are identified. *Message authentication and integrity*, to protect messages from alteration and allow receivers to corroborate the node that created the message. If necessary, *entity authentication* can provide evidence of the sender *liveness*, that is, the fact it generated a message recently. To prevent a sender from denying having sent a message, *non repudiation* is needed. Furthermore, *access control* and *authorization* can determine what each node is allowed to do in the network, in terms of the implemented system functionality. *Confidentiality* can keep message content secret from unauthorized nodes.

Related to information hiding, *privacy and anonymity* are required, at least at the level of protection achieved before the advent of VC systems. In general, VC systems should not disclose or allow inferences on private user information. In particular, the identity of a vehicle performing a VC-specific action (e.g., transmitting a message) should be concealed. Anonymity, with respect to an observer, depends on the set of vehicles: an observer cannot determine which among all vehicles in that set performed an action. In other words, any two actions by the same vehicle cannot be linked. Under specific circumstances, an observer could consider a vehicle more likely to perform an action.

Rather than seeking strong anonymity along with authentication (and other security requirements), less stringent requirements are considered. Cryptographically protected messages should not allow the identification of their sender, and two or more messages generated by the same vehicle should be difficult to link to each other. More precisely, messages produced by a node over a protocol-selectable period of time $\tau$ can be linked, but messages $m_1, m_2$ generated at times $t_1, t_2$ such that $t_2 > t_1 + \tau$ cannot. The shorter $\tau$ is, the fewer the linkable messages are, and the harder to trace a node becomes.

Beyond security and anonymity, *availability* is also sought, so that VC systems remain operational even in the presence of faults, and resume normal operation after the removal of the faulty nodes. Another significant dimension is that of *non-cryptographic security*, including the determination of data *correctness* or *consistency*. Traditionally, if the sender of a message is trusted, then the content of the message is trusted as well. This notion is valid for long-lived, static trust relationships, but in VC systems there is often no ground for similar approaches. It is thus necessary to assess the *trustworthiness of data per se*, as obtained by other nodes in the VC system.

### C. Basic Secure VC System Elements

VC systems will rely on multiple *Certification Authorities* (CAs), each managing the *long-term* identities and credentials for nodes registered in its *region* (e.g., canton or state). Each node is uniquely identified, and it holds one or more private-public key pairs and certificates. The CA attests to the attributes of each registered node, according to its capabilities and roles in the system. The CAs are responsible for *evicting* nodes from the system, for administrative or technical reasons, if needed. The CAs interact infrequently with nodes, utilizing RSUs as a gateway.

The basic tool for nodes to *secure communication* is to digitally sign messages, after attaching a time-stamp and the signer's location and certificate to the message. This way, modification, forgery, replay, and relay attacks can be defeated. The latter relates to secure neighbor discovery [12], which is possible exactly because safety beacons include the time and location at the point they are sent across the wireless medium [13], [14]. Signatures can be applied in different ways, to beacons, multi-hop flooded, and position-based [3] multi- or uni-casted messages, not only by the message originator but also by relaying nodes.

To provide both security and a degree of anonymity, long-term keys and credentials are *not* used to secure communication. Rather, the approach of *pseudonymity* or *pseudonymous authentication* is used. Each vehicle is equipped with multiple certified public keys (pseudonyms) that do not reveal the node identity. It obtains those pseudonyms by a trusted third party, a *pseudonym provider* (PNP), by proving it is registered with a CA. Then, the vehicle uses each pseudonym and private key for at most for $\tau$ seconds, the pseudonym lifetime, and then discards it. Messages signed under the same pseudonym can be trivially linked, but messages signed under different pseudonyms cannot.

## III. ON DESIGN CHOICES AND APPROACHES

### A. To Secure or Not to Secure?

One may pose a legitimate question: *Is an arguably complex secure VC system necessary*? Cellular telephony and nomadic wireless Internet access were deployed without strong security features, they proliferated and continue to do so, in spite of significant security and privacy breaches. But the significant differences of VC systems from those two systems imply why a different approach is necessary. First, cellular and nomadic network access rely on infrastructure, which simplifies the provision of security; for example, associations (trust) is needed between the mobile node and the infrastructure. Moreover, a compromise would not have costly or even fatal consequences, as those of a multi-car accident caused by attacking a VC system.

With higher stakes for VC systems (reducing accidents, saving lives, improving transportation), a single security incident would perhaps suffice for the public to loose confidence in this new technology. Then, assuming sufficient security is in place, one can ponder on the *degree of protection* a security architecture should and can offer. Would strong cryptographic protection of network and application protocols suffice to ensure that false data are not injected in the system? Or, would anonymous or pseudonymous authentication ensure location privacy when numerous cameras and optical plate recognition systems are deployed?

VC security does not address problems that are present independently of the use of VC. But it fends off a broad range of exploits that could otherwise wreak havoc to VC and the transportation system. External adversaries or vehicle modifications are mitigated, the results of key compromise are thwarted, and accountability, even if anonymous authentication is used, can lead to the eviction of adversarial nodes. Until this happens, redundancy or absence of corroborating evidence by other near-by vehicles could enable inferring the truthfulness of received VC application messages. Initial results are promising, showing that *data centric trust establishment* is feasible [16]. Investigations for specific applications, complex environments, as well as measures to thwart determined adversaries, can lead to stronger protection.

### B. Public Key Cryptography or What?

The high volatility and large scale of VC systems led to the choice of digital signatures. For the choice of appropriate algorithms the following basic factors were considered: The processing times for signature generation and verification, the security overhead (public key and signature sizes), the standardization of cryptographic algorithms (and confidence in their strength), and the experience in implementation. Elliptic Curve-based algorithms (e.g. EC-DSA) seem to be preferred, primarily because of low network overhead for strong security.

Security levels for VC system entities have not been concretized yet, but 80-bit or higher security seem to be favored, to prevent practical cryptanalytic attacks. It is important though to consider *for which operation* a specific security level for cryptographic primitives is needed. Clearly, CAs and PNPs should have higher security levels. Then, high security would be needed for long-term keys and certificates. The lowest level should be assigned to short-term keys, for which security levels below 80-bits could be perhaps considered. Even if such a key, valid for minutes or hours, were broken within weeks or months by a determined adversary, there would be no immediate consequences, only a reduction in overhead. Of course, this would be true if and only if VC traffic were not used in the long-run, for example, logged for future liability attribution (Sec. IV-C).

We can also consider alternatives to "classic" public key cryptography, which can lead to usability limitations if strong anonymity is sought. The larger the number of pseudonyms a vehicle needs (in the extreme, one per message it transmits), the higher the overhead for generating the key pairs and certificates will be. Obtaining those would be costly, if done over a cellular link for example. But privacy protection or even availability problems can appear if vehicles are unable to obtain new pseudonyms. The use of anonymous long-term cryptographic material and *anonymous authentication* can enable on-board, on-the-fly generation of short-term keys and credentials. This is shown to be a viable approach in [1], which can be strengthened to prevent abuse of anonymous authentication [2].

### C. Are Secure VC Systems Practical?

With specific proposals in the literature, it is important to assess their practicality: *Can secure VC systems satisfy stringent application requirements*? This is a fundamental question for VC systems in general, but it becomes all the more relevant due to the significant communication and processing overhead and other restrictions security and privacy-enhancing mechanisms impose. Consider, for example, a safety application that notifies on emergency braking: can vehicle collisions be still avoided after adding security? Simulation studies over a range of settings and protocols for pseudonymous authentication indicate that with the appropriate system design, secure VC can be in practice as effective as unsecured VC [1], [7].

*D. Processing Power*

The OBU processing power is critical for the practicality of secure VC systems. The cryptographic validation of received messages is the primary component of processing load, especially in dense topologies. Based on results in [1], [7], also considering anonymous authentication as part of a pseudonymous authentication system, experimental platforms currently considered for VC systems would be insufficient. They could not sustain cryptographic processing, even if revocation status checking (and thus overhead) were not considered, under dense network conditions (e.g., in congested multi-lane highways). Beyond optimizations, *increased processing power* is needed, with cryptographic co-processors being one option, so that nodes validate all or a high fraction of messages within application constraints (e.g., in the order of 100ms for safety).

*E. Can VC Equipment be Trusted?*

The low physical protection of VC system nodes and the motivation of attackers raise the question if VC equipment can be trusted. It appears that if this were the case, with the appropriate security in place, the overall problem of securing VC systems could be addressed. However, cost is major concern, and making the entire on-board equipment tamper-proof or tamper-resistant would be impractical. Nonetheless, there are critical resources that must be protected: a *Trusted Component (TC)*, for example, as proposed by SeVeCom, should store private keys and perform private cryptographic operations. With a tamper-resistant TC, extraction of the private keys would be impossible. With a real-time clock and battery integrated in the TC, the adversary would be unable to feed the TC with fake future time-stamps and obtain falsified cryptographically protected messages.

*F. Is Revocation Necessary?*

To ensure the robustness of the VC system, it is important to *evict* faulty nodes and prevent the utilization of compromised keys. The distribution of *Revocation Lists* (RLs) is the basic approach to do this in VC systems [6], [11], complemented by other defense mechanisms or enhancements [5], [15]. More important, leveraging on relatively sparsely placed road-side infrastructure, and with low bandwidth consumption by RSUs, all vehicles can obtain the latest RL within an average commute time [11]. Nonetheless, the processing overhead to control if a node is in the RL can be high, primarily because of anonymity mechanisms. For "classic" cryptography, each pseudonymous certificate should be validated at first reception, but with many pseudonyms per vehicle the RL will be large. For anonymous credentials, each received message should be checked against the much shorter RL, but each check is orders of magnitude costlier than that for "classic" cryptography.

The challenge is not the RL distribution but rather its on-board processing cost, which is proportional to the RL length. The question follows naturally: *Is it necessary to ignore messages signed by all nodes in a RL?* In fact, this is closely related to the composition of the RL. For example, if a stolen vehicle is in the RL, its VC equipment is not necessarily compromised; thus, it would be unwise for the safety of receiving vehicles to ignore its messages. A flexible approach to address the problem could reduce the length of the RL and thus the processing overhead: Distinct RLs are created, according to the "urgency" of using them for real-time message validation. The RL of highest priority, processed at all times, can contain only evidently faulty or compromised nodes. At a lower priority can be a RL with nodes that are potentially faulty, perhaps, in different RLs according to the type of fault, and at the lowest priority an RL with nodes evicted for other reasons. Lower priority RLs can checked if possible or if a specific event triggers the need to do so (e.g., suspected faulty behavior by a near-by node).

## IV. SOME NARROW PASSAGES AHEAD

*A. Bringing VC to the Market*

The way VC systems will be deployed can significantly influence security solutions and thus the system trustworthiness. The primary question is whether the VC deployment will be *monolithic* or *evolutionary*, with no negative or positive connotation for either term. In other words, will the OBU be one or two powerful, multi-purpose box(es), or perhaps a multi-core processor, running all protocols? Or, will it be a set of boxes, each of them added on-board gradually, running a single application with just enough processing power for the specific tasks?

The monolithic model resembles what is considered thus far in the development of secure VC architectures. But the evolutionary approach may be closer to what a strongly market-centric deployment commands, driven by the applications (e.g., entertainment) preferred by consumers. Reflecting the mind-set of some stakeholders, the evolutionary deployment would lead most likely to a minimum application-specific security, as well as a heterogeneous on-board network. The situation would become more involved if user devices (e.g., PDAs, cell-phones, home or corporate computers) interact with the OBUs, to obtain for example useful personal information for navigation, record trip data into a personal log, or access physical spaces or digital content. All these aspects would raise new challenges in terms of security and privacy.

*B. Organizational Issues*

The reliance on authorities (Sec. II) is in sync with long-lived approaches in managing vehicles [10]. However, the efforts to operate CAs often result in a degree of skepticism, with frequently recurring questions on the CA operational cost or the difficulties of collaboration among diverse CAs. The alternative of vehicle manufacturers running their own CAs is considered. Nonetheless, this raises concerns on monopoly or oligopoly situations that could be imposed this way, or even the likelihood that proprietary solutions that do not provide full-fledged security might be adopted.

Existing *multi-domain* systems such as cellular systems, which require access control and accounting, indicate that addressing organizational issues is feasible. In fact, the success story of cellular systems can provide useful clues. Numerous distinct providers, each having high numbers of registered clients and devices, each of them uniquely identified and able to operate in other regions and billed for network usage via its "home" provider, provide interesting features and even similarities to ponder on.

*C. Legal Issues*

*User awareness* of the offered protection is paramount: The guarantees the VC system offers, the residual vulnerabilities, and the role of all system entities should be clearly stated to end user agreements. Analogies can be drawn to existing systems, for example, with recent privacy breaches against cellular telephony perpetrated via mobile device exploits or by insiders. The responsibility of each entity, the users included, should be clear, as this also relates to VC equipment maintenance and accreditation.

The use of VC systems towards assisting the attribution of *liability* for transportation incidents is a controversial issue. Clearly, a non-secure VC system would be out of the question for such a task. Strong accountability in secure VC systems, as discussed above, is possible. But determining which entity and under which circumstances can perform this, and then through which procedure liability can be attributed, is far from straightforward.

Policies for VC systems will also have to deal with the issue of *voluntary or mandatory* use of the equipment. Will, for example, safety and traffic efficiency functionality be mandatory, the same way nowadays seat-belts are? It is likely that users have the incentive, for example, to run those applications in order to lower their insurance premiums. But if deployment is mandatory, would privacy concerns be fully addressed? Solutions discussed above can indeed achieve this. But users may still raise legitimate arguments in favor of powering off their VC boxes. Or, perhaps raise the need for distinct secure VC instantiations, for example, for government vehicles that do not wish to take any risk of being traced by terrorists.

## V. WHAT DOES THE FUTURE HOLD?

Significant progress has been made already towards comprehensive security solutions for VC systems. It is very interesting to realize that all these results are produced at a stage when the actual VC systems are several years from deployment. Essentially, *security for VC systems is being developed at design time*. This allows for a deep and early understanding of the problem at hand. Equally important, security designs can influence VC protocols and applications. For example, OBU characteristics can be set to a certain standard to enable security; or protocol features that enhance performance but allow high-impact exploits can be disabled for enhanced resilience.

Of course, as the VC protocols and applications mature, and continue to evolve after deployment, an even clearer understanding of threats and possible exploits will be possible. In all cases, it is paramount to avoid misconceptions on that front, with a characteristic example Global Navigation Satellite Systems (GNSS) used to provide accurate location and time. With few exceptions, commercial systems such as GPS are assumed secure, even though attacks even against upcoming systems allow an adversary to manipulate the position estimated by GNSS receivers [9]. If VC systems are deployed with meagre or no defenses, high-profile or disastrous exploits can take place, leading at a later stage authorities and car manufacturers to summon security experts to address the problem. But with the current extensive interest, rising awareness, and significant results, there is a unique opportunity to have trustworthy VC systems at their initial deployment.

### REFERENCES

[1] G. Calandriello, P. Papadimitratos, J.-P. Hubaux, and A. Lioy. Efficient and robust pseudonymous authentication in vanet. In *ACM VANET*, Montreal, Quebec, Canada, 2007.
[2] J. Camenisch, S. Hohenberger, M. Kohlweiss, A. Lysyanskaya, and M. Meyerovich. How to win the clone wars: Efficient periodic n-times anonymous authentication. In *ACM Computer and Communication Security (CCS)*, Alexandria, VA, October 2006.
[3] C. Harsch, A. Festag, and P. Papadimitratos. Position-based routing for vanets. In *IEEE VTC Fall 2007, Baltimore, MD, USA*, Oct. 2007.
[4] IEEE1609.2. IEEE trial-use standard for wireless access in vehicular environments - security services for applications and management messages, July 2006.
[5] T. Moore, M. Raya, J. Clulow, P. Papadimitratos, R. Anderson, and J-P. Hubaux. Fast exclusion of errant devices from vehicular networks. In *IEEE SECON*, San Francisco, CA, USA, June 2008.
[6] P. Papadimitratos, L. Buttyan, J-P. Hubaux, F. Kargl, A. Kung, and M. Raya. Architecture for secure and private vehicular communications. In *ITST'07*, Sophia Antipolis, France, 2007.
[7] P. Papadimitratos, G. Calandriello, A. Lioy, and J-P. Hubaux. Impact of vehicular communication security on transportation safety. In *IEEE MOVE 2008, in conjunction with INFOCOM*, Phoenix, Arizona, USA, April 2008.
[8] P. Papadimitratos, V. Gligor, and J.-P. Hubaux. Securing vehicular communications - assumptions, requirements, and principles. In *Workshop on Embedded Security in Cars (ESCAR)*, Berlin, Germany, November 2006.
[9] P. Papadimitratos and A. Jovanovic. Protection and fundamental vulnerability of gnss. In *International Workshop on Satellite and Space Communications (IWSSC)*, Toulouse, France, October 2008.
[10] P. Papadimitratos, A. Kung, J.-P. Hubaux, and F. Kargl. Privacy and identity management for vehicular communication systems: A position paper. In *Workshop on Standards for Privacy in User-Centric Identity Management*, Zurich, Switzerland, July 2006.
[11] P. Papadimitratos, G. Mezzour, and J.-P. Hubaux. Certificate revocation list distribution in vehicular communication systems (short paper). In *ACM VANET 2008*, San Francisco, CA, September 2008.
[12] P. Papadimitratos, M. Poturalski, P. Schaller, P. Lafourcade, D. Basin, S. Čapkun, and J.-P. Hubaux. Secure neighborhood discovery: A fundamental element for mobile ad hoc networking. *IEEE Communications Magazine*, February 2008.
[13] M. Poturalksi, P. Papadimitratos, and J.-P. Hubaux. Secure neighbor discovery in wireless networks: Formal investigation of possibility. In *ACM ASIACCS*, March.
[14] M. Poturalksi, P. Papadimitratos, and J.-P. Hubaux. Towards provable secure neighbor discovery in wireless networks. In *ACM Workshop on Formal Methods in Security Engineering*, October.
[15] M. Raya, P. Papadimitratos, I. Aad, D. Jungels, and J.-P. Hubaux. *IEEE Journal on Selected Areas in Communications, Special Issue on Vehicular Networks*, February 2007.
[16] M. Raya, P. Papadimitratos, V. Gligor, and J.-P. Hubaux. On datacentric trust establishment in ephemeral ad hoc networks. In *IEEE INFOCOM*, Phoenix, Arizona, USA, April 2008.